\documentstyle[12pt,aasms4]{article}

\slugcomment{Not to appear in Nonlearned J., 45.}

\lefthead{Djorgovski et al.}
\righthead{Collapsed Cores in Globular Clusters}

\begin{document}

\title{Invariance violation extends \\
the cosmic ray horizon ?}

\author{Tadashi Kifune}
\affil{Institute for Cosmic Ray Research, University of Tokyo,
    Tanashi, Tokyo 188-8502, Japan}

\begin{abstract}
We postulate 
that the energy-momentum relation is modified 
for very high energy particles  
to violate Lorentz invariance and  
the speed of photon is changed from the light velocity $c$.  
The violation effect is amplified, in a sensitive way to detection, 
through the modified kinematical constraints on the conservation of 
energy and momentum,  
in the absorption process     
of $\gamma$-rays colliding against photons of longer wavelengths   
and converting into an electron-positron pair.  
For $\gamma$-rays of energies higher than $10^{13}$ eV,   
the minimum energy of the soft photons for the reaction 
and then 
 the  absorption mean free path of $\gamma$-rays   
are altered  by orders of magnitude 
from the ones conventionally estimated.  
Consideration is similarly applied to 
 high energy cosmic ray protons.  
The consequences may require the standard assumptions on
 the maximum distance that very high energy radiation can travel from
to be revised.
\end{abstract}


\section{Introduction}

Phenomena strikingly deviating from 
the conventional understanding can take place   
if the speed of photons   
is allowed to vary from the constant light velocity $c$.   
Coleman and Glashow (1997) argued that 
the velocity $c_0$ of light in the vacuum in a frame moving relative 
to the rest frame of the universe 
can differ from the maximal attainable speed 
$c$ of a material body. 
If $c_0 > c$, the photon 4-momentum $(\vec E /c_0, E)$ is time-like, 
allowing a sufficiently energetic photon to decay 
into an electron-positron pair. 
Alternatively, if $c_0 <c$, 
a charged particle will lose 
energy via vacuum \v Cerenkov radiation    
because the particle speed approaching $c$ at high energies 
can exceed the light speed $c_0$. 
Thus,  the distance that energetic $\gamma$-rays or  
 charged particles can travel through  
is shortened for either $c_0 > c$ or $c_0 < c$. 
However, 
the  violation of relativity may also have a possibility of 
increasing the travelling distance of high energy radiation 
 rather than decreasing.  

The speed of photons can vary  
due to modifications of the vacuum (Latorre et al. 1995). 
Different energies excite the vacuum differently, and 
with quantum gravity, 
Amelino-Camelia et al. (1997) speculated that 
the energy-momentum relation for photons  
is modified to violate Lorentz invariance and      
the anomalous effect increases 
as photon energy approaches  
the Planck energy scale ($ \sim 10^{19}$ GeV).       
Amelino-Camelia et al. (1998)  has proposed 
a test for 
the modified speed by measuring   
the arrival time structure of photons 
from objects at cosmological distances.     
The propagation speed  
is 
$dE / dk = c (1 - \xi _{\gamma} E / E_0)$  
from the modified relation 
of energy $E$ and momentum $k$,  
\begin{equation}
  c^2 k^2 = (E + \xi _{\gamma} {E^2 \over 2 E_0})^2 
          = E^2 + \xi _{\gamma} {E^3 \over  E_0}  
\end{equation}
  with $\xi _{\gamma} = \pm 1$ and 
$E_o \sim 10^{19}$ GeV, 
according to Amelino-Camelia et al. 1998. 
By assuming the existence of a preferred frame for 
Lorenz transformation but preserving other 
kinds of invariance,  
 Coleman and Glashow 1998 gives 
 a relation, which is similar to (1) and equivalent to  
$c^2 k^2 = E^2 - m_i^2 c^4+ (c^2-c_i^2) E^2$ 
for $ck \approx E \gg m_i c^2$, where     
$c_i$ is the maximal attainable speed of particle species $i$ 
with mass $m_i$. 
Our study in what follows will be in an explicit form based on (1), 
but 
the results  are, to some extent,  
applicable also to the other case  
by putting $c^2-c_i^2$ in the place where 
the term $\xi _i E/E_0$ appears. 

The electromagnetic radiation of the shortest wavelength   
so far detected is TeV $\gamma$-rays 
(see for example, Weekes et al. 1997; Ong 1998), 
but
even for these high energy photons, the effect of velocity variation 
remains as small as a fraction $\sim 10^{-15}$ of 
the light velocity.      
However, a higher degree of anomalous effect is expected    
for the absorption process, 
in which $\gamma$-rays are  converted 
into an electron--positron  pair
when they collide  with soft photons of the universal radiation field.  
 Photons of TeV energies react with infrared photons,   
and PeV $\gamma$-rays with the 2.7K CMB (cosmic microwave background).   
  From the number density of the photons of 
the universal infrared and CMB, 
  the collision mean free path is usually estimated as 
  about 100 Mpc and 10 kpc ({\it e.g.} Strong et al. 1974; 
Stecker et al. 1992), respectively 
for TeV and PeV $\gamma$-rays.  
The largest distance we can see with the high energy radiations    
is thus much shorter than 
those of  X-rays and optical 
photons of longer wavelengths. 

\section{Kinematical constraints on the photon-photon collision process}

The threshold energy is determined  by  
 a function of the total energy $W$ and momentum $Q$  
 of the initial state, $W^2 - c^2Q^2$, 
which must exceed $(2 mc^2)^2$ for the pair creation, 
where $m$ is the electron mass.   
In the initial state of a $\gamma$-ray obeying the relation (1) 
and soft photons of energy $\varepsilon$, 
the square of the energy $E$ and the momentum $\sim E/c$ in  
$W^2 - c^2Q^2$ 
almost cancel each other to leave 
 $4\varepsilon E - \xi _{\gamma} E^3/E_0$   
(the two photons are assumed to be from opposite directions).     
The remaining two terms are, 
when $\varepsilon / E$ is as small as $\xi_{\gamma} E / E_0$, 
 of similar magnitude to each other,  
incurring a considerable degree of anomalous effect.   
In addition,  
the two terms become as large as $m^2 c^4$ when 
$E \approx 10^{12} \xi_{\gamma} ^{-1/3} \approx 10^{12} \varepsilon ^{-1}$ eV, 
and the threshold condition is considerably affected by the violation effect  
for $E > 10^{13} \xi_{\gamma} ^{-1/3}$ eV, which can be much lower 
than the Planck energy scale. 

Let us examine the above speculation by calculating in our rest frame 
 the conservation of momentum of the reaction.  
By putting  $c$ = 1 hereafter, 
the momentum $E - \varepsilon + \xi _{\gamma} E^2 /(2E_0)$ 
of the initial state 
(we assume for simplicity, but not to lose the generality, 
$\lq$head-on' collision of $\gamma$-ray and 
soft photon)    
 must be equal to 
$p_1 cos \alpha + p_2 cos \beta$ of 
 the final state, where 
($E_1$, $\vec p_1$) and ($E_2$, $\vec p_2$) is energy and momentum  
of electron and positron and $E + \varepsilon = E_1 + E_2$. 
The $\alpha$ and $\beta$ are the angles $\vec p_1$ and $\vec p_2$ 
make against the direction of the initial gamma ray 
and are in the order of $m / E_i$ ($i = 1$ and 2).
By using the approximation $p_i = E_i - m^2/(2E_i)$ and 
the condition on the transverse momenta   
$\alpha p_1 = \beta p_2$, we obtain 
\begin{equation}  
 -2 \varepsilon + {\xi _{\gamma} E^2 \over 2 E_0} = 
- {E + \varepsilon \over 2} \cdot 
( {m^2 \over E_1 E_2} + \alpha \beta ).  
\end{equation}
We use a parameter $x$ to replace $E_1$ and $E_2$ 
by $E_{1, \, 2} = (1 \pm x) (E + \varepsilon )/2$ 
with the condition  $| x | \le 1 - 2m /(E + \varepsilon )$.    
The equation (2) is then written as 
\begin{equation}
\varepsilon \approx 
  {m^2 \over E} \cdot {1  \over 1 - x^2 } + 
  {\alpha \beta E \over 4}    
 + \xi _{\gamma} \cdot {E^2 \over 4 E_0}.     
\end{equation}
\noindent 
The first and second terms are of the same order of 
magnitude, the latter term  
approaching zero at the threshold of the reaction.  
With increasing $E$, softer photons can serve as 
the {\it target} photons,   
shortening the absorption mean free path. 
This conventional view is altered 
as $E$ further increases, by the presence of 
the term $\xi _{\gamma} E^2 / (4 E_0)$.  
However, electron and positron may also obey a
 modified relation similar to (1). 
Let us put, similarly to (1), 
the square of the modified momentum $q$ to be equal to 
\begin{equation}
q^2 = p^2 + \xi _e {E^3 \over E_0} 
       = E^2 - m^2 + \xi _e {E^3 \over E_0}.    
\end{equation} 
\noindent 
We grant $|\xi _e | \ne 1$ for allowing electrons to have 
a different  degree of  
the invariance violation.  
By replacing $p$ by $q$, 
a relation similar to (3), 
\begin{equation}
\varepsilon 
= {m^2 \over E} \cdot {1  \over 1 - x^2 } 
  + {\alpha \beta E \over 4} 
+  {E^2 \over 4 E_0} (\xi _{\gamma} - \xi _e \cdot {1 + x^2 \over 2 }),    
\end{equation} 
\noindent is obtained. 
  Since energy $E$ is divided 
 almost equally into two particles (when $x \approx 0$) 
and the anomalous term has $E^2$ dependence, 
 the factor 2 appears in the denominator of  $\xi _e (1 + x^2)/2$, 
thus reducing the effect in the final state. 
The fact might seem to  suggest that  
the single $\lq$leading' particle 
taking almost all the total energy ($|x| \approx 1$) is capable of 
eliminating the anomalous effect. 
However, the substitution of  $|x| = 1 - 2m /(E + \varepsilon )$ 
  into (5)  
increases the energy $\varepsilon$ 
 to  $m \sim 1$ MeV which is much higher  than $m^2 /E$, 
and then leads to 
a very large travelling distance,   
resulting in  no less anomalous consequence. 
Only if $\xi _e$ is tuned to be $2\xi _{\gamma}$ for unknown reason, 
we can recover the ordinary relation for $\varepsilon$. 

\section{Comparison with observations}

Energy $\varepsilon$ of the {\it target} photon is plotted 
against $\gamma$-ray energy $E$ in Fig.~1.   
A minimum value of  $\varepsilon$  is taken,  
if $\xi = 2 \xi _{\gamma} - \xi_e > 0$, at the critical energy 
$E_c = (4m^2 E_0 /\xi )^{1/3} \approx 10^{13}$ eV 
 for $\xi  = 1$. 
The minimum value is about 0.03 eV and 
 $\varepsilon$ must increase to about 30 eV at $E = 10^{15}$ eV.  
Thus, CMB of $\sim 10^{-3}$ eV is never given a chance    
of contributing  to the pair creation process.  
The travelling length has the shortest value of $\sim$10 Mpc 
at $E = E_c$, and increases again for $E > E_c$.  
Since $\sim 30$eV photons are very sparse, 
$\gamma$-rays would  travel through  distances $>$ 1 Gpc. 
When $\xi <0$, 
$\varepsilon$ must rapidly decrease as $E$ approaches $E_c$.  
Before taking negative value,   
 CMB photons become available as  {\it target}.   
The energetic $\gamma$-rays are then efficiently absorbed 
with a mean free path of $\sim 10$ kpc, which is shorter   
 by orders of magnitude than the ordinary case. 

\placefigure{fig1}

Among the TeV $\gamma$-ray sources are   
two  active galaxies, Mrk 421 and Mrk 501 (Punch et al. 1992; 
Aharoninan et al. 1997; 
Catanese et al. 1997).  
The objects are at distances of $\sim$ 100 Mpc, 
and the detection is consistent   
with the absorption mean free path determined by 
infrared intensity in the extra-galactic space. 
The energy spectrum of the $\gamma$-rays is found to extend 
 up to 10 - 20 TeV.  
If the spectrum continues to extend to even higher energies, 
 $\xi < 0$ would be excluded.  
Because, if $\xi < 0$,  a very sharp cut-off 
at the common energy $E_c$ is expected in the energy spectrum 
for any extra-galactic objects.  
The negative $\varepsilon$ for $E > E_c$ in the case of $\xi < 0$ 
implies that 
$\gamma$-rays 
does not need {\it target} photons for pair creation. 
The threshold of $\gamma$-ray    
decaying into  an electron-positron pair 
(Coleman and Glashow 1997) 
is given by $|E_c|$.  
By putting $|E_c| = -E_c = (4m^2 E_0/(-\xi ))^{1/3} > 20$ TeV, 
 we obtain a limit on $\xi$ to be $\xi > -1.3$ 
(by putting $E_0 = 1 \times 10^{28}$ eV).
On the other hand, when $\xi$ is positive and $\sim 1$, 
the travelling distance 
turns over to increase above about 10 TeV and 
to become larger than 1 Gpc for the  photons 
around 100 TeV.  
 Thus, 100 TeV photons would arrive  
from extra-galactic sources,  
 implying that evidence of detection would be strongly 
 in favour of the $\xi >0$ case.  
Extra-galactic diffuse $\gamma$-rays are 
most likely from distant unresolved active galaxies. 
The energy spectrum is observed by Sreekumar et al. 1997  
to extend to about 50 GeV 
with a hard power index of $\sim -2.3$.    
 The spectrum is expected to become steeper at higher energies 
due to the intenser infrared photons and then CMB at longer wavelengths. 
However, in the case of $\xi >0$,  
the energy spectrum may have a {\it turn-over} 
of showing a harder slope above $E = E_c$. 
Electrons lose their energy quickly and have   
difficulty in producing such high energy $\gamma$-rays. 
However,  
cosmic ray protons are known to have energy spectrum extending 
to $10^{20}$ eV, and  
 are able to produce  $\gamma$-rays as high as or larger than 100 TeV. 
The increase of travelling distance suggests more 
contribution from more distant sources and  thus we may expect 
 considerably intense, diffuse $\gamma$-rays 
in the  higher energy region.
  From the Crab nebula, 50 TeV photons have been detected 
(Tanimori et al. 1997) by using  imaging \v Cerenkov 
telescope at large zenith angles. 
Detection of 100 TeV $\gamma$-rays is not unrealistic 
by the extensive application of such methods. 

\section{Case of cosmic ray protons}

Our phenomenological consideration can be extended 
to other processes of soft photons acting as the reaction target. 
Extremely high energy cosmic ray protons 
({\it e.g.} Hill and Schramm 1985; 
Axford 1994; O'Halloran et al. 1998; 
Takeda et al. 1998)
 are believed to lose energy 
by pion production in collision with CMB.   
The number density of CMB and the photo-pion cross section 
determine the collision mean free path to be $\sim 10$ Mpc. 
The energy spectrum thus suffers from the GZK cutoff 
(Greisen 1966; Zatsepin and Kuzmin 1966) 
at about $10^{20}$ eV,     
if the protons are from the distances farther than the mean free path. 
However, the relation (4) assumed for proton and pion alters   
 the threshold of the energy $\varepsilon$ of {\it target} photons 
from $\mu M /(2E)$ to   
\begin{equation}
\varepsilon 
= {\mu M \over 2E} +  \Xi \cdot {E^2 \over 4 E_0} , 
  \ \ {\rm where } \ \ \Xi =  \xi _p 
   -  \xi _p ({M \over M + \mu })^2 -  
    \xi _{\pi} ({\mu \over M + \mu })^2.  
\end{equation}
\noindent 
The energy $E$ is for the initial proton, 
$M$ and $\mu$ are the proton and pion mass, 
and $\xi _p$ and $\xi _{\pi}$ for protons and pions, respectively.  
We have put the parameter $x$ to be $(M -\mu) / (M+\mu)$,  
 corresponding to the case in which the ratio of proton to pion energy  
in the final state is $(1+x)/(1-x) = M / \mu$  
near the threshold of the process.  
If $\Xi$ is positive and not very far from the order of one, 
the second term becomes comparable with $ \mu M /(2E)$  
at the critical energy  
$E_{p\pi} = ( M \mu E_0 /\Xi)^{1/3} \approx 10^{15}$ eV. 
At $E \approx 10^{20}$ eV,   
the energy $\varepsilon$ needs to be, as dominated by $\Xi E^2 /(4E_0)$,  
 as large as $10^{12}$ eV,  
 and CMB  can not play the role of {\it target} photons. 
Cosmic ray protons at $\sim 10^{20}$ eV 
can then travel without the attenuation  
 from cosmological distances farther than $\sim 10$ Mpc. 
As a result, the  energy spectrum of cosmic rays would extend 
 beyond $10^{20}$ eV,  
provided that 
the  acceleration to such high energies takes place 
in celestial objects.    
There is speculation
by Vietri 1995 and Waxman 1995 
 that GRB ($\gamma$-ray bursts) provides   
the origin of the highest energy cosmic rays.  
If so, GRBs at cosmological distances 
would produce a cosmic ray spectrum without GZK cutoff.   
If $\Xi < 0$, 
$\varepsilon$ can be $\sim 10^{-3}$ eV 
when $E$ approaches $|E_{p\pi}| = - E_{p\pi} \approx 10^{15}$ eV, 
implying that  
 the {\it target} photons of CMB gives 
  a characteristic  attenuation length of 
$\sim$10 Mpc to $10^{15}$ eV cosmic rays. 
 For $E > |E_{p\pi}|$, 
the kinematical condition does not require the {\it target} photon  
for the pion production to occur and  
the processes of     
 $p \rightarrow p + \pi^0$ and $p \rightarrow n + \pi^+$  
will cause protons to rapidly lose energy.    
The energy loss due to the spontaneous production of pion is 
in contradiction with 
the evidence of cosmic ray protons up to $10^{20}$ eV.    
Thus, we  limit $|E_{p\pi}|$ longer than $10^{20}$ eV, and  
 obtain a bound of $\Xi > - 10^{-17}$. 

\section{Discussions}

Collision processes of high energy radiations with soft photons 
 enable us to test the consequences of 
violation of  Lorentz invariance.   
 In particular, the effect of quantum gravity 
which Amelino-Camelia et al. (1998) suggest  
can be made detectable by the reaction process in the energy region 
much lower than the Planck energy scale,  but 
in the sufficiently high energy region 
to which the validity of Lorentz transformation  has not been  
examined yet by accelerator experiments.   
The violation suggests an interesting possibility of  
increasing  the distance we can reach  
by observing  high energy $\gamma$-rays and cosmic rays. 
Absence of the GZK cutoff would not necessarily 
mean nearby origin of the extremely high energy cosmic rays. 
Photons of energies larger than 100 TeV from extra-galactic objects 
would provide a clear test because it is hard to explain  by  
 other mechanisms than the violation of invariance.  
It seems impossible, once we admit the relation (1) for photons, 
to totally eliminate 
the anomalous effect,  without a specially designed   
tuning among   
 the invariance violating terms for various particles of different masses.   
Other ways have been proposed to alter the conventional view
on the relativity (Gonzares-Mestres 1997; Sato 1998).
 Absence of GZK cutoff can be explained, according to Sato 1998, 
without abandoning the Lorenz invariance, but 
by giving a special status to $\lq\lq$the universe frame''
in which CMB is isotropic.

Protons are expected to lose  energy  
via vacuum \v Cerenkov radiation (Coleman and Glashow 1997),  
when $\xi _{\gamma} -\xi _p >0$. 
The threshold energy,  
 $E_{th} = (M^2 E_0 / (\xi _{\gamma} - \xi _p))^{1/3}$, 
of the radiation is calculated from  
 equating the speed of photon to that of proton.  
 The detection of the highest energy $10^{20}$ eV of cosmic rays 
 sets a constraint  $E_ {th} > 10^{20}$ eV   
and limits the $\xi$-parameters to be  
$\xi _{\gamma} -\xi _p < 10^{-15}$ 
 or rather $\xi _{\gamma} \le \xi _p$  for practical use.
The \v Cerenkov radiation from electron yields 
a less tight constraint  
because of the dependence of $\xi$ on $E_{th}$, {\it i.e.}   
$\xi = \xi_{\gamma} - \xi_e \propto E_{th}^{-3}$.   
The highest energy $2 \times 10^{12}$ eV  
of cosmic ray electrons ever detected 
 (Nishiumura et al. 1997) must be less than 
$E_{th}$ and then we obtain 
$\xi _{\gamma} -\xi _e < 10^3$.  
If we choose, for the bound on $E_{th}$,  
$\sim 100$ TeV electrons 
which are responsible for TeV $\gamma$-rays
and which are considered to extend over a spatial scale of 
at least $\sim$ 1 pc, 
the  constraint is made somewhat tighter 
as $\xi _{\gamma} -\xi _e < 10^{-2}$.    
We may argue about allowed region of $(\xi _{\gamma}, \xi_e)$ values 
by combining the condition $-1.3  < 2\xi _{\gamma} - \xi _e$ 
 from the decay process 
$\gamma \rightarrow e^+ + e^-$. 
The region is, to some extent, in disfavor with negative values. 
The time structure of the outbursts of TeV $\gamma$-rays 
from Mrk~421 and Mrk~501 sets constraint  
on $\xi _{\gamma}$.   
The delay time of 10 TeV $\gamma$-rays relative to 
 photons of lower energies  is  
$(d/c)\xi_{\gamma} (10^{13} {\rm eV} /E_0) \approx 10 \xi_{\gamma} 
\, s$ for distance  $d=100$ Mpc. 
Duration of the outbursts which has been 
observed typically as a few hours 
is considered to be longer than the delay time, 
 providing  a constraint of $\xi_{\gamma} < 10^3$. 
We need to know the outburst  structure  
 in time scales as short as $\sim 10 s$ 
to constrain $|\xi|$ less than $\sim 1$. 
In order to restrict the $|\xi|$ parameter more tightly,    
 It is very useful to observe  
photons of higher energies from more distant objects.  
However, we can not expect that such $\gamma$-rays 
 survive against 
conversion into an electron-positron pair and reach us,     
unless we premise that the Lorenz invariance is violated. 
  
We may argue that 
 the extensive air shower events of $\sim 10^{20}$ eV  
are in truth not due to proton.  
Such a claim is relevant to $\lq$correct' understanding of  
the complex phenomena of the cascade shower process,  
which still leaves a room for a speculation 
of $\lq$anomalous interaction'  of cosmic rays.  
Thus, it is interesting to indicate that 
the modified relation of energy and momentum can affect detection 
 of the high energy radiations. 
Observation of $\gamma$-rays for example is based on the 
 pair creation process    
in the detector material (satellite experiments) or in the atmosphere 
(ground-based observation for TeV photons and cosmic rays).  
The recoil momentum $\delta$ of atomic nucleus in the material  
is calculated as  
twice the right hand side of the expression (5). 
The ordinary recoil momentum is  
$\delta _0 \approx m^2 E/(2E_1 E_2) = 0.1$ eV/c 
 at $E \approx 10^{13}$ eV, 
which corresponds to quite a small energy transfer 
to the recoiled atomic nucleus of a large mass. 
When $\xi _{\gamma} =1$ and $E > E_c$, 
$\delta$ must be larger than $\delta _0$ 
but remains reasonably small so that the process would 
not be largely affected. 
When $\xi _{\gamma} = - 1$ and $E > E_c$, 
 $\delta$ appears negative, the recoil being 
 in the opposite direction to the initial $\gamma$-ray. 
There are generally several processes which compete 
the pair creation and for which 
the invariance violation is of different strengths.  
The consequence may appear as observation of an anomalous behaviour   
of very high energy radiations. 

\acknowledgments

The author thanks the  anonymous referee for the comments and advice  
that have brought the revised form and content of the present paper.     


\clearpage

\figcaption[fig.eps]{
Energy of {\it target} photons 
against $\gamma$-ray energy. 
The number density of  
{\it target} photons decreases with their energy $\varepsilon$ and thus   
 the mean free path generally increases with $\varepsilon$. 
The travelling distance of $\gamma$-rays is 
indicated on the right hand side of the vertical axis. 
The thin line (1) shows the ordinary case of 
$\varepsilon = m^2 / E$, and 
the dashed line (2) represents the anomalous contribution 
$\xi E^3 /(4 E_0)$, where $\xi = 2\xi _{\gamma}- \xi _e$ 
as shown in the expression (5) of the text. 
The result of adding the both terms 
is shown as indicated by the thick lines for $\xi =1$ (a) 
and $-1$ (b) cases. 
In the model of Coleman and Glashow 1998,  
the  line (a) has a less steep increase with energy because of the 
factor $c^2-c_i^2$ instead of $\xi _i E/E_0$.   
\label{fig1}}


\begin{thebibliography}{}

\bibitem[Aharonian et al.\ 1997]{aha97} Aharonian, F.A. et al. 1997, 
     Astron. Ap., 327, L5
\bibitem[Amelino-Camelia et al.\ 1997]{ame97} 
     Amelino-Camelia, G., Ellis, J., Mavromatos, N.E., and 
     Nanopoulos, D.V., 1997, Int. J. Modern Phys. A, 12, 607
\bibitem[Amelino-Camelia et al.\ 1998]{ame98}
     Amelino-Camelia, G., Ellis, J., Mavromatos, N.E.,  
     Nanopoulos, D.V., and Sarkar, S., 1998, Nature, 393, 763
\bibitem[Axford 1994]{axf94} 
     Axford, W.I., 1994, \apjs, 90, 937
\bibitem[Catanese et al.\ 1997]{cat97} 
     Catanese, M. et al.\ 1997, \apjl,  487, L143
\bibitem[Coleman et al.\ 1997]{col97} 
     Coleman, S. and Glashow, S.L., 1997, Phys. Lett., B 405, 249 
\bibitem[Coleman et al.\ 1998]{col98} 
     Coleman, S. and Glashow, S.L., 1998, hep-ph 9812418 
\bibitem[Gonzalez-Mestres 1997]{gon97} 
     Gonzalez-Mestres, L., 1997, Proc. 25th Int. Cos. Ray Conf., 
     6, 113
\bibitem[Greisen 1966]{gre66} 
     Greisen, K., 1966, Phys. Rev.  Lett., 16, 748 
\bibitem[Hill and Schramm 1985]{hil85} 
     Hill, C.T. and Schramm, D.N., 1985, Phys. Rev. D,  
     31, 564
\bibitem[Latorre et al.\ 1995]{lat95} 
     Latorre, J.I., Pascual, P., and Tarrach, R., 1995,  
     Nucl. Phys., B437, 60
\bibitem[Nishimura et al. 1997]{nis97} 
     Nishimura, J. et al., 1997, Proc. 25th Int. Cos. Ray Conf., 
     4, 233
\bibitem[O'Halloran et al.\ 1998]{oha98} 
     O'Halloran, T., Sokolsky, and Yoshida, S., 1998,  
   Physics Today, January, 31
\bibitem[Ong 1998]{ong98} 
     Ong, R.A., 1998, Phys. Rep., 305, 95 
\bibitem[Punch et al.\ 1992]{pun92}
     Punch, E. et al., 1992, Nature, 358, 477
\bibitem[Sato 1998]{sat98} 
      Sato, H., 1998,  
     Black holes and high energy astrophysics 
    (edited by H. Sato and N. Sugiyama: Universal Academy Press), 
     401
\bibitem[Sleekumar et al.\ 1997]{sre97} 
      Sreekumar, P., Stecker, F.W., and Kappadath, S.C., 1997,  
      AIP Conf. Symp. Proc. (4th Compton Symp.) 410, 344
\bibitem[Stecker et al.\ 1992]{ste92} 
     Stecker, F.W., De Jager, O.C., and Salamon, M.H.,  
     1992, \apjl, 390, L49
\bibitem[Strong et al.\ 1974]{str74} 
      Strong, A.W., Wdowczyk, J. and Wolfendale, A.W., 
      1974, J. Phys. A, 7, 1767 
\bibitem[Takeda et al.\ 1998]{tak98} 
     Takeda, M. et al., 1998, Phys. Rev. Lett., 81, 1163
\bibitem[Tanimori et al.\ 1997]{tan97} 
      Tanimori, T. et al., 1997, \apjl, 492, L33
\bibitem[Vietri 1995]{vie95} 
     Vietri, M., 1995, \apjl, 453, 883
\bibitem[Waxman 1995]{wax95} 
    Waxman, E., 1995, Phys. Rev. Lett., 75, 386 
\bibitem[Weekes et al.\ 1997]{wee97} 
     Weekes, T.C., Aharonian, F.A., Fegan, D.J., and Kifune, T., 
     1997, AIP Conf. Symp. Proc. (4th Compton Symp.), 410, 361
\bibitem[Zatsepin and Kuzmin 1966]{zat66} 
         Zatsepin, G.T. and Kuzmin, V.A., 1966, Soviet Phys. JETP Lett.,  
      4, 78
\end{thebibliography}
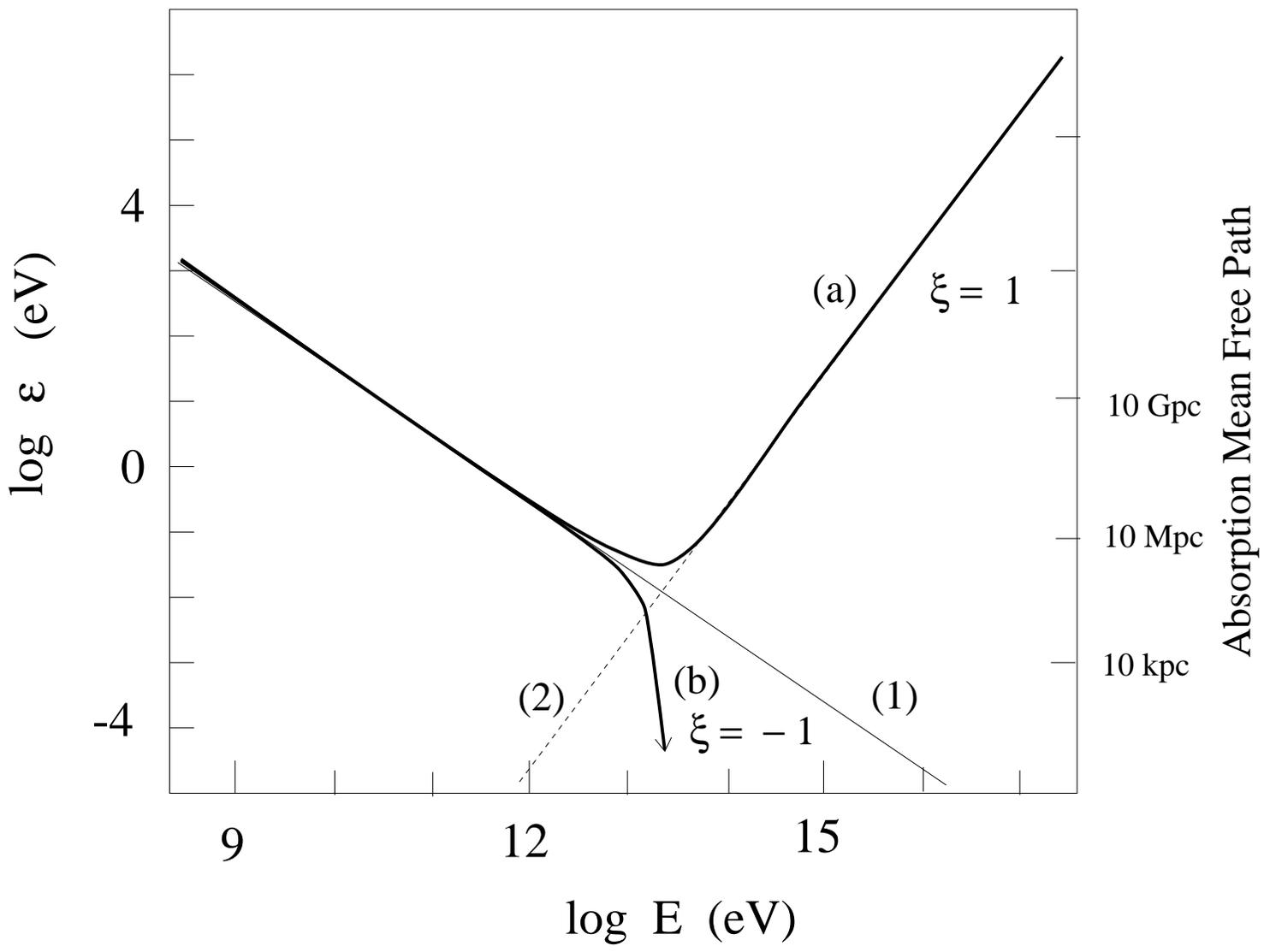
\end{document}